\documentclass[seceq,preprint]{ptptex}

\usepackage{graphicx,bbm}
\usepackage{hyperref}

\preprintnumber[3cm]{
YITP-09-46
}

\newcommand{\id}{{\mathbbm 1}}

\def\slash#1{\not\!#1}

\title{
Lattice Fermions Based on Higher-Dimensional Hyperdiamond Lattices%
}

\author{
Taro \textsc{Kimura}$^{1}$%
and Tatsuhiro \textsc{Misumi}$^{2}$}

\inst{
$^{1}$Department of Basic Science, University of Tokyo, Tokyo 153-8902, Japan\\
$^{2}$Yukawa Institute for Theoretical Physics, Kyoto University, Kyoto 606-8502, Japan\\
}

\abst{
In this paper we generalize to higher dimensions several types of fermion actions on the hyperdiamond lattice including a two-parameter class of minimal-doubling fermions ``Creutz fermion'' and a simple fermion with sufficient discrete symmetry ``BBTW fermion''.
Then it is shown that they possess some properties in common with the four-dimensional case:
BBTW fermions in higher even dimensions inevitably yield unphysical degrees of freedom.
Creutz fermions are defined on the distorted lattices, and they lose the high discrete symmetry of the original lattices.
We also find properties specific to the higher-dimensional cases.
The parameter range for Creutz action to yield minimal-doubling and physical fermions becomes narrower with the dimension getting higher, thus it becomes more and more difficult to realize minimal-doubling.
In addition, we generalize the subspecies of Creutz and BBTW actions including a new class of minimal-doubling actions ``Appended Creutz action''.
}

\begin{document}

\maketitle

\section{Introduction}\label{sec:intro}
 
The relativistic electron system on a two-dimensional honeycomb lattice, graphene\cite{neto:109}, has been attracting a great deal of attention for these several years.
From the viewpoint of lattice field theory,
the electron in this system is regarded as a fermion appropriate for the lattice simulation since it has some desirable properties such as locality, chiral symmetry, the minimal number of fermion doublings and sufficient discrete symmetry for a good continuum limit.
Among them, the minimal fermion doubling is the most outstanding characteristic in this fermion.
Although there exist only two (or three) light quarks in QCD, 
one fermion with chiral symmetry and other common features on a four-dimensional lattice yields  massless fermions of multiple number of two in a continuum limit as a result of Nielsen-Ninomiya's no-go theorem\cite{Nielsen:1980rz,Nielsen:1981xu,Nielsen:1981hk}. 
In addition, the lattice fermions which bypass the no-go theorem such as domain-wall fermion\cite{Kaplan:1992bt,Furman:1994ky} and overlap fermion\cite{Neuberger:1998wv} satisfying Ginsparg-Wilson relation\cite{Ginsparg:1981bj} demand an expensive numerical task.
On the other hand, as the two light quarks, chirality of the
minimal-doubling two fermions have opposite signs.
Therefore the chirally symmetric fermion including only minimal number of doublers (two fermions) will be much faster and more useful in the lattice simulation since the two fermion degrees of freedom can be directly interpreted as the two light quarks in QCD.
In the past, Wilczek \cite{Wilczek:1987kw} and Karsten \cite{Karsten:1981gd} proposed the fermion action with the minimal number of doublings called ``Karsten-Wilczek fermion''.
However, since this action lacks sufficient discrete symmetry to prohibit relevant and marginal operators to be generated through quantum corrections, one need fine-tuned parameters to take a good continuum limit.
Therefore, successful construction of the chirally symmetric fermion both with minimal-doubling and the requisite discrete symmetry has been longed for.

Recently, there have been some attempts to generalize the graphene, which possesses both the minimal doublers and the sufficient discrete symmetry, to a four-dimensional system and apply it to the lattice QCD simulation.
In these attempts they try to construct Dirac fermion on a four-dimensional honeycomb lattice, called hyperdiamond lattice, with keeping the desirable properties. 

The naive Dirac fermion on a hyperdiamond lattice was constructed by Bedaque et. al.\cite{Bedaque:2008jm} by defining left and right-handed fermions on the two sublattices respectively.
This ``Bedaque-Bachoff-Tiburzi-WalkerLoud (BBTW) fermion'' action includes only the nearest hopping terms and has the sufficient discrete symmetry, although it yields more than the minimal number of doublings.
In our recent work\cite{KM:2009co} it is pointed out that this fermion action inevitably produces unphysical poles of fermion propagator since the BBTW Dirac operator independently includes $i\gamma_{\mu}$-terms and $\gamma_{\mu}\gamma_{5}$-terms corresponding to vector and axial-vector functions.

The most successful Dirac fermion on the hyperdiamond lattice is ``Creutz fermion'' with the two parameters ($B, C$), which had been proposed by Creutz\cite{Creutz:2007af} before BBTW fermion was presented.
What is notable about Creutz fermion is that it exhibits the minimal amount of fermion doublers.
However it is pointed out in Ref. \citen{KM:2009co,Bedaque:2008xs} that this action lacks sufficient discrete symmetry to prohibit the redundant operators like Karsten-Wilczek fermion.\cite{Capitani:2009yn} Thus it has no advantage in the lattice simulation compared to the conventional lattice fermions, although it still attracts our interest because of its exotic properties.

The purpose of this paper is to generalize the fermion actions on the hyperdiamond lattice including BBTW and Creutz actions to higher dimensions.
It is quite non-trivial if we can construct fermions on the higher-dimensional hyperdiamond lattice since it has non-trivial lattice structure unlike the hypercubic lattice. 
Besides, the general-dimensional viewpoint leads to deeper understanding of these fermions.
In this paper we successfully construct the several types of fermions on general-dimensional hyperdiamond lattices.
Then we find that BBTW fermion inevitably yields unphysical poles of fermion propagator also in higher dimensions.
On the other hand Creutz fermions in higher dimensions are defined on distorted hyperdiamond lattices as long as one keeps physicality of the fermion poles, and the action loses the high discrete symmetry of the original lattice as the four-dimensional case.
It is pointed out that the range of the parameters for Creutz action to yield minimal-number and physical fermions becomes narrower with the dimension getting higher.
Thus, in higher dimensions it becomes more and more difficult to realize the minimal-doubling Creutz fermion.
We also derive the higher-dimensional versions of the special parameter conditions, one of which is called ``Creutz condition'' in which the fermion poles form an exact hyperdiamond lattice in momentum space, and the other of which is called ``Bori\c{c}i condition'' in which the lattice becomes orthogonal.

In our recent paper\cite{KM:2009co} the subspecies of Creutz and BBTW actions are discussed.
In this paper we also generalize these subspecies.
And it is shown that all of them produce unphysical fermion degrees of freedom.
In addition we generalize to higher dimensions a new class of minimal-doubling actions ``Appended Creutz action'' proposed in \citen{KM:2009co}.
Then it is pointed out that this fermion has the deformed momentum-space structure which leads to the modified Creutz and Bori\c{c}i conditions compared to the original Creutz action.

In Sec.\ref{sec:hyperdiamond} we study the hyperdiamond lattices in general dimensions.
In Sec.\ref{sec:BBTW} we generalize BBTW fermion to higher even dimensions and study the properties.
In Sec.\ref{sec:creutz} we generalize Creutz fermion and investigate the
characteristics, and also study the subspecies of Creutz and Bedaque actions in general even dimensions.
In Sec.\ref{sec:appended} we show a higher-dimensional version of Appended Creutz action.
Sec.\ref{sec:summary} is devoted to a summary and discussion.

\section{Hyperdiamond lattice}\label{sec:hyperdiamond}

In this section, we formulate general-dimensional hyperdiamond lattices.
To construct higher-dimensional lattices, we start with some geometries.
When we denote a $d$-dimensional object, $d+1$ points are required.
For example, a line can be described when two points are given.
In the cases of the diamond lattice $(d=3)$ and the honeycomb lattice $(d=2)$, each site
has four and three isotropic bonds, respectively.
Therefore, they realize the isotropic ``minimal-bond lattices'' in each
dimensional space.
Generalizing this property, we define $d$-dimensional hyperdiamond lattice with $d+1$
vectors $\{ \rm{e}^\alpha \}$ satisfying
\begin{equation}
 \rm{e}^\mu \cdot \rm{e}^\nu = \left\{ \begin{array}{ccc}
			    1 & \mbox{for} & \mu = \nu \\ \cos \theta_d &
			     \mbox{for} & \mu \not= \nu
				   \end{array}\right. . \label{bond_vec01}
\end{equation}
Due to the isotropic condition of the lattice, the bond vectors sum up to zero,
\begin{equation}
 \sum_{\mu=1}^{d+1} \rm{e}^\mu=0. \label{sumuptozero}
\end{equation}
And the inner products between these vectors can be obtained by
(\ref{sumuptozero}) as
\begin{equation}
 \cos \theta_d = -1/d.
\end{equation}  
Then we present an explicit expression of $d$-dimensional bond vectors satisfying these conditions,
\begin{equation}
 \begin{array}{lccrrrrrrrc}
 \rm{e}^1 & = & (& c_1 s_2 \cdots s_{d-1} s_d ,& c_2 s_3 \cdots s_d ,& \cdots & , & c_{d-2} s_{d-1} s_d, & c_{d-1} s_d,& c_d &) \\
 \rm{e}^2 & = & (& s_2 \cdots s_{d-1} s_d ,& c_2 s_3 \cdots s_d ,& \cdots & , & c_{d-2} s_{d-1} s_d, & c_{d-1} s_d,& c_d &) \\
 \rm{e}^3 & = & (& 0 ,& s_3 \cdots s_d ,& \cdots & , & c_{d-2} s_{d-1} s_d, & c_{d-1} s_d,& c_d &) \\
  &\vdots &&&&&&&& \\
 \rm{e}^{d-1} & = & (& 0,& 0,& \cdots & , & s_{d-1} s_d , & c_{d-1} s_d,& c_d & ) \\  
 \rm{e}^{d} & = & (& 0,& 0,& \cdots & , & 0, & s_d ,& c_d & ) \\  
 \rm{e}^{d+1} & = & (& 0,& 0,& \cdots & , & 0, & 0 ,& 1 & ) \\  
 \end{array} \label{bond_vec}
\end{equation}
with $ c_\mu \equiv \cos \theta_\mu = -1/\mu$, $s_\mu \equiv \sin \theta_\mu = \sqrt{\mu^2-1}/\mu$.
Another expression of the four-dimensional hyperdiamond lattice is presented in \citen{Bedaque:2008jm,KM:2009co}.

\begin{figure}[tbp]
\begin{center}
  \includegraphics[width=11.5em]{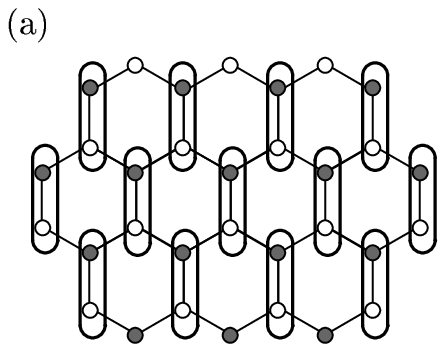}\quad
  \includegraphics[width=11.5em]{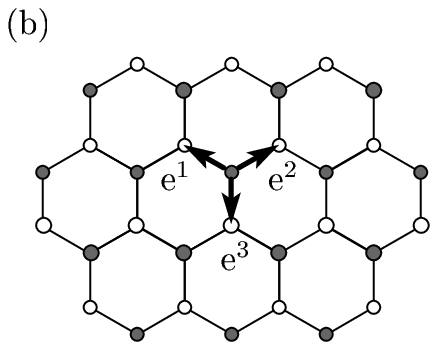}\quad
  \includegraphics[width=11.5em]{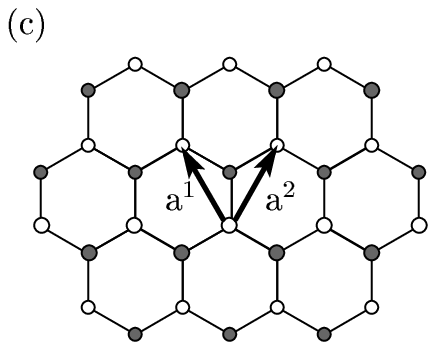}
\end{center}
 \caption{Two dimensional diamond (honeycomb) lattice. (a) unit cells
 are encircled and consist of two kinds of sites, shaded $A$ and
 open $B$ sites. (b) bond vectors $\left\{{\rm e}^\mu\right\}$ from $A$ to $B$ sites and
 (c) primitive vectors $\left\{{\rm a}^\mu\right\}$ characterizing translation symmetry of the lattice.}
 \label{honeycomb_lattice}
\end{figure}

Next, let us define primitive vectors $\{\rm{a}^\mu\}$ which characterize
the translation invariance of the lattice.
As shown in Fig.\ref{honeycomb_lattice}, an unit cell of the hyperdiamond lattice has two sites, called $A$
and $B$ site.
Specifying one of the bond vectors ${\rm e}^{d+1}$ as the vector from
$A$ to $B$ in the same unit cell, primitive vectors become
\begin{equation}
 {\rm a}^\mu = {\rm e}^\mu - {\rm e}^{d+1} \quad \mbox{for} \quad
  \mu=1,\cdots,d. \label{primitive_vec}
\end{equation}
When the position of an $A$ site is denoted as
$x=\sum_{\mu=1}^d x_\mu \rm{a}^\mu$, the corresponding $B$ site is
on the site of $x+{\rm e}^{d+1}$. 
Therefore, $(d+1)$-th vector ${\rm e}^{d+1}$ transforms $A$ and $B$ sites
to each other in the same unit cell. 
This fact is similar to the Clifford algebra, and ${\rm e}^{d+1} =
-\sum_{\mu=1}^d \mathrm{e}^\mu$
corresponds to $\gamma^{d+1} \propto \gamma^1 \cdots \gamma^d$ which generates
the chiral transformation.

Here it is shown that the angle between the primitive vectors does not
depend on the dimensionality of the lattice,
\begin{equation}
 \cos \eta = \frac{\mathrm{a}^\mu \cdot \mathrm{a}^\nu}{\left|\mathrm{a}^\mu\right|
  \left|\mathrm{a}^\nu\right|} = \frac{1}{2} \quad \mbox{for} \quad \mu\not= \nu.
\end{equation}
This reads the angle $\eta = \pi/3$, and thus the equilateral triangular
structure is observed in any dimensions.
The shortest length between the sites of the same sublattice is of course given by
$\left|\mathrm{a}^\mu\right|=\left|\mathrm{e}^\mu-\mathrm{e}^{d+1}\right|$. 
Then, $\mathrm{a}^\mu-\mathrm{a}^\nu=\mathrm{e}^\mu-\mathrm{e}^\nu$ also
gives the shortest length.
Thus the number of the nearest unit cells is $d(d+1)$, and
the Wigner-Seitz cell consists of $d(d+1)$ hyperplanes.
This unit cell structure will play an important role in section
\ref{sec:appended} on considering some lattice actions.

On the other hand, reciprocal vectors $\{\mathrm{b}_\mu\}$, basis for the
momentum space are obtained as 
\begin{equation}
 \mathrm{b}_\mu = \frac{d}{d+1}\mathrm{e}^\mu \quad \mbox{for} \quad \mu=1,\cdots,d
\end{equation}
satisfying $\mathrm{a}^\mu \cdot \mathrm{b}_\nu = \delta^\mu_\nu$. 
An arbitrary momentum vector $p$ is represented by non-orthogonal basis, 
\begin{equation}
 p = \sum_{\mu=1}^d \left( p \cdot\mathrm{a}^\mu\right) \mathrm{b}_\mu.
\end{equation}
As the primitive vectors, let us find the nearest reciprocal sites.
The unit reciprocal vector gives the shortest length between the reciprocal sites,
$\left|\mathrm{b}_\mu\right|=d/(d+1)$. 
This length is also obtained by
$\left|\mathrm{b}_1+\cdots+\mathrm{b}_d\right|=d/(d+1)$. 
Then, the number of the nearest reciprocal sites is $2(d+1)$ and the
Brillouin zone can consist of them.

\section{Higher-dimensional BBTW action}\label{sec:BBTW}

We now introduce higher-dimensional actions on the hyperdiamond lattices
defined in the previous section. 
In this section, we consider higher-dimensional generalizations of BBTW
action proposed in \citen{Bedaque:2008jm}. 
BBTW action is a naive lattice action preserving the symmetry of the exact
hyperdiamond lattice.

Let us start with Clifford algebras to consider spinor structures of
lattice fermions.
We choose a representation of $2m$-dimensional Clifford algebra as
\begin{equation}
 \Gamma^{(2m)}_\mu = \tau_1 \otimes \Gamma^{(2m-2)}_\mu
 = \left( \begin{array}{cc}
    0 & \Gamma^{(2m-2)}_\mu \\ \Gamma^{(2m-2)}_\mu & 0
	  \end{array} \right) \quad \mbox{for}
  \quad \mu = 1, \cdots, 2m-1, \qquad 
\end{equation}
\begin{equation}
 \Gamma^{(2m)}_{2m} = \tau_2 \otimes \id_{\left[2^{m-1}\right]}
 = \left( \begin{array}{cc}
   0 & -i \id_{[2^{m-1}]} \\ i \id_{[2^{m-1}]} & 0 
	  \end{array} \right),
  \qquad
 \Gamma^{(2m)}_{2m+1} = \tau_3 \otimes \id_{\left[2^{m-1}\right]}
 = \left( \begin{array}{cc}
    \id_{[2^{m-1}]} & 0  \\ 0 & -\id_{[2^{m-1}]}
	  \end{array}\right),
\end{equation}
where $\tau$'s are Pauli matrices acting on the sublattice structure of
the hyperdiamond lattice, and $\id_{[n]}$ is the $n$-component
identity matrix. 
These gamma matrices $\Gamma$'s satisfy anti-commutation relations, $\{
\Gamma^{(2m)}_\mu, \Gamma^{(2m)}_\nu \} = 2 \delta_{\mu\nu} \id_{[2^m]} $.
For convenience, we denote $\Gamma^{(2m)}$ as a $2m$-dimensional vector,
\begin{equation}
 \Gamma^{(2m)} = \left( \begin{array}{cc}
		  0 & \bar{\gamma}^{(2m)} \\
		 \gamma^{(2m)} & 0
			\end{array} \right)
\end{equation}
with
\begin{equation}
 \gamma^{(2m)} = \left( \Gamma^{(2m-2)}_1, \cdots,
		  \Gamma^{(2m-2)}_{2m-1}, i \id_{[2^{m-1}]} \right),
 \quad
 \bar{\gamma}^{(2m)} = \left( \Gamma^{(2m-2)}_1, \cdots,
		  \Gamma^{(2m-2)}_{2m-1}, -i \id_{[2^{m-1}]} \right).
 \label{spinor_vec}
\end{equation}
As discussed in the previous section, the hyperdiamond lattice consists
of two sublattices.
Such a sublattice structure can be interpreted as an internal degree of freedom, e.g. staggered fermion\cite{Susskind:1976jm}.
To construct Dirac fermion on $2m$-dimensional
hyperdiamond lattice, we introduce $2^{m-1}$-component fermions  $\psi^A$, $\psi^B$ on $A$ and $B$ sites.
In the case of the hyperdiamond lattice, they are interpreted as chiral, Weyl fermions.
Here we notice that $\psi^A$, $\bar{\psi}^B$ are left-handed, $\bar{\psi}^A$,
$\psi^B$ are right-handed, and $\psi^A$ and $\bar{\psi}^A$
are not hermite conjugate, but independent degrees of freedom.

We then consider lattice actions on the hyperdiamond lattices with
nearest neighbor hoppings, which we call the higher-dimensional BBTW action.
Because an $A$ site is surrounded by only $B$ sites, and vice versa,
nearest neighbor hoppings correspond to off-diagonal parts of Dirac operator.
In odd dimensional cases, we can not choose all of space-time gamma
matrices as off-diagonal, and thus a lattice action can not be naively
constructed with only nearest neighbor hopping terms.
Thus we introduce $2m$-dimensional BBTW action
\begin{eqnarray}
  S_{\mathrm{BBTW}} & = & \sum_x \Big[ \sum_{\mu=1}^{2m} \left( \bar{\psi}^A_{x-\mathrm{a}_\mu} \left( \gamma^{(2m)} \cdot \mathrm{e}^\mu \right) \psi^B_x 
	- \bar{\psi}^B_{x+\mathrm{a}_\mu} \left( \bar{\gamma}^{(2m)} \cdot \mathrm{e}^\mu \right) \psi^A_x \right) \nonumber \\
	&& \qquad + \bar{\psi}^A_{x} \left( \gamma^{(2m)} \cdot \mathrm{e}^{2m+1} \right) \psi^B_x
	- \bar{\psi}^B_{x} \left( \bar{\gamma}^{(2m)} \cdot \mathrm{e}^{2m+1} \right) \psi^A_x \Big].
 \label{BBTW_action}
\end{eqnarray}
We note that, to construct the lattice action, $\mathrm{e}_\mu$ should
not be the bond vector, and thus the primitive vector should not be
associated with them by the relation
$\mathrm{a}_\mu=\mathrm{e}_\mu-\mathrm{e}_5$.

In momentum space,%
\footnote{
We use non-orthogonal coordinates such as $p_\mu = p \cdot \mathrm{a}^\mu$.
}
this action is rewritten as
\begin{equation}
 S_{\mathrm{BBTW}} = \int \frac{d^{2m} p}{(2\pi)^{2m}} \ \bar{\Psi}(-p) D(p) \Psi(p)
\end{equation}
with 
\begin{equation}
 \Psi(p) = \left( \begin{array}{c} \psi^A(p) \\ \psi^B(p) \end{array} \right), \quad
 \bar{\Psi}(p) = \left( \bar{\psi}^A(p) \quad \bar{\psi}^B(p) \right), 
 \label{Dirac_spinor}
\end{equation}
and the associated Dirac operator becomes
\begin{equation}
 D(p) = i \sum_{\mu=1}^{2m} \left( \Gamma^{(2m)} \cdot \mathrm{e}^\mu \right) \sin p_\mu
 + \Gamma^{(2m)}_{2m+1} \Gamma^{(2m)} \cdot \left( \sum_{\mu=1}^{2m} \mathrm{e}^\mu \cos p_\mu + \mathrm{e}^{2m+1} \right).\label{Dirac_op1}
\end{equation}
We can also write down the alternative representation of the Dirac
operator such that the chiral symmetry is easy to see as following,
\begin{eqnarray}
D(p)&=&
 \left( \begin{array}{cc}
0 & z(p) \\
-z^{\dag}(p) & 0
	\end{array}\right),
\label{chiralrep}
\\
z(p)&=&(\gamma^{(2m)}\cdot{\rm e}^{2m+1})1+(\gamma^{(2m)}\cdot{\rm e}^{2m})e^{ip_{2m}}+\cdots
+(\gamma^{(2m)}\cdot{\rm e}^{1})e^{ip_{1}}.
\label{BBTW_chiral}
\end{eqnarray}
Then, one can confirm the chiral invariance of this Dirac operator by looking into the relation with $\Gamma^{(2m)}_{d+1}$ as 
\begin{equation}
\Gamma^{(2m)}_{2m+1} D(p) \Gamma^{(2m)}_{2m+1}=D^{\dag}(p)=-D(p).
\end{equation}
To investigate fermionic excitations in the lattice action, we expand
this operator with two kinds of chiral, anti-hermitian basis,
$i\Gamma^{(2m)}$ and $\Gamma^{(2m)}_{2m+1}\Gamma^{(2m)}$ as
\begin{equation}
 D(p) = i \sum_{\mu=1}^{2m} x_\mu(p) \Gamma^{(2m)}_\mu +
  \sum_{\mu=1}^{2m} y_\mu(p) \Gamma^{(2m)}_{2m+1} \Gamma^{(2m)}_\mu, \label{Dirac_op2}
\end{equation}
where explicit coefficients for the choice of the expression of the hyperdiamond lattice (\ref{bond_vec}) are represented as
\begin{equation}
 x_\mu(p) = \left( \prod_{\nu=\mu+1}^{2m} s_\nu \right) \left( c_\mu \sum_{\nu=1}^\mu \sin p_\nu + \sin p_{\mu+1} \right)
 \quad \mbox{for}\quad \mu = 1, \cdots, 2m-1,
\end{equation}
\begin{equation}
 y_\mu(p) = \left( \prod_{\nu=\mu+1}^{2m} s_\nu \right) \left( c_\mu \sum_{\nu=1}^\mu \cos p_\nu + \cos p_{\mu+1} \right) 
 \quad \mbox{for}\quad \mu = 1, \cdots, 2m-1,
\end{equation}
\begin{equation}
 x_{2m}(p) = c_{2m} \sum_{\nu=1}^{2m} \sin p_\nu, \qquad
 y_{2m}(p) = c_{2m} \sum_{\nu=1}^{2m} \cos p_\nu +1.
\end{equation}

According to BBTW's paper\cite{Bedaque:2008jm}, we now search spectral
zeros of Dirac operator (\ref{Dirac_op1}) of the form $p =
\tilde{p}_\mathrm{B} (\sigma_1, \cdots, \sigma_{2m})$ with $\sigma_\mu
= \mathrm{sgn}(p_\mu)$, and we assume that $\#\{\sigma_\mu= +1\} =
\#\{\sigma_\mu= -1\} = m$.
In this case, some coefficients vanish, $x_{2m}(p) =
y_{\mu}(p)=0\ (\mu=1, \cdots, 2m-1)$, and the operator
(\ref{Dirac_op2}) reduces to
\begin{equation}
 D(p) = i \tau_1 \otimes \left( \sum_{\mu=1}^{2m-1} x_\mu(p)
		       \Gamma^{(2m-2)}_\mu - y_{2m}(p)\id_{[2^{m-1}]} \right).
\end{equation}
Thus, four eigenvalues of Dirac operator are given by
\begin{equation}
 \pm i \left( \pm \sqrt{ \sum_{\mu=1}^{2m-1}x_\mu^2(p) } + y_{2m}(p) \right),
\end{equation}
and the condition that Dirac operator has zero modes is
$\sum_{\mu=1}^{2m-1}x_\mu^2(p) = y_{2m}^2(p)$.
Solving this condition, we obtain poles of BBTW Dirac operator,
\begin{equation}
 \cos \tilde{p}_{\mathrm{B}} = 1, \quad -\frac{m}{m+1}.
\end{equation}
The former kind of poles are degenerated at $p_\mu=0$ and the number of
poles is one, but the number of the latter kind of poles is 
$\left( \begin{array}{c} 2m \\ m \end{array} \right) = (2m)!/(m!)^2$.
In the four-dimensional case, a pole at $p_\mu=0$
and non-zero valued poles located on $p_1 = - p_2 = - p_3 = p_4 =
\cos^{-1}\left(-2/3\right)$, etc. are obtained, and they are consistent
with our result\cite{Bedaque:2008jm}.

Indeed this action is chiral invariant as shown in (\ref{BBTW_chiral}).
However, as discussed in our recent paper\cite{KM:2009co}, BBTW
Dirac operator includes not only $i\Gamma$-terms but $\Gamma
\Gamma_{d+1}$-terms corresponding to vector and axial-vector functions,
i.e. BBTW Dirac operator (\ref{Dirac_op1}) is rewritten as
\begin{equation}
 D(p) = \sum_{\mu=1}^{2m} \left[ i\left(\Gamma^{(2m)}\cdot
			    \mathrm{e}^\mu\right) q_\mu \cos
 \tilde{p}_{\mathrm{B}}
 - \left(\Gamma^{(2m)}_{2m+1} \Gamma^{(2m)} \cdot
    \mathrm{e}^{\mu}\right) \sigma_\mu q_\mu \sin \tilde{p}_{\mathrm{B}} 
			    \right] + {\cal O}(q^2)
\end{equation}
with expanded momentum around a pole $p_\mu = \tilde{p}_{\mathrm{B}}
\sigma_\mu + q_\mu$. 
This means that we can not apply Nielsen-Ninomiya's no-go
theorem\cite{Nielsen:1980rz,Nielsen:1981xu,Nielsen:1981hk} to
this kind of lattice actions, and there exist no guarantees for the
number of doublers and the covariance of poles.
In fact, non-zero valued poles can not be reduced to covariant Dirac
form, which correspond to unphysical, or mutilated fermions\cite{Celmaster:1982ht,Celmaster:1983jq,Drouffe:1983kq}.
Thus, we claim that BBTW fermions describes unphysical degrees of freedom also in higher dimensions.

\section{Higher-dimensional Creutz action}\label{sec:creutz}

In this section, we generalize the minimal-doubling action proposed by
Creutz\cite{Creutz:2007af}.
Creutz action was directly constructed in momentum
space, then the spatial configuration of the action was shown in
\citen{Bedaque:2008jm}. 
Creutz action is based on the hyperdiamond lattice deformed by two
parameters, and includes non-nearest neighbor hopping terms.

According to the above discussion, we deform a higher-dimensional
hyperdiamond lattice (\ref{bond_vec}) by two parameters, $B$ and $C$, as the
four-dimensional case,%
\footnote{
Our notation is different from the four-dimensional
case\cite{Creutz:2007af,Bedaque:2008jm}: $B$ $\to$ $\sqrt{d+1}B$.
} 
\begin{equation}
 \begin{array}{lccrrrrrrrc}
 \mathrm{e}^1 & = & (& c_1 s_2 \cdots s_{d-1} s_d ,& c_2 s_3 \cdots s_d ,& \cdots & , & c_{d-2} s_{d-1} s_d, & c_{d-1} s_d,& Bc_d &) \\
 \mathrm{e}^2 & = & (& s_2 \cdots s_{d-1} s_d ,& c_2 s_3 \cdots s_d ,& \cdots & , & c_{d-2} s_{d-1} s_d, & c_{d-1} s_d,& Bc_d &) \\
 \mathrm{e}^3 & = & (& 0 ,& s_3 \cdots s_d ,& \cdots & , & c_{d-2} s_{d-1} s_d, & c_{d-1} s_d,& Bc_d &) \\
  &\vdots &&&&&&&& \\
 \mathrm{e}^{d-1} & = & (& 0,& 0,& \cdots & , & s_{d-1} s_d , & c_{d-1} s_d,& Bc_d & ) \\  
 \mathrm{e}^{d} & = & (& 0,& 0,& \cdots & , & 0, & s_d ,& Bc_d & ) \\  
 \mathrm{e}^{d+1} & = & (& 0,& 0,& \cdots & , & 0, & 0 ,& BC & ) \\   \end{array} \label{bond_dist}
\end{equation}
where we note that these bond vectors reduces to these of the exact
hyperdiamond lattice by choosing the particular parameters: $B=C=1$.
Then we construct Creutz action on $2m$-dimensional distorted
hyperdiamond lattice
\begin{eqnarray}
  S_{\mathrm{C}} & = & \frac{1}{2} \sum_x \Big[\sum_{\mu=1}^{2m}  \Big(
  \bar{\psi}_{x-\mathrm{a}_\mu}^A \left( \Sigma^{(2m)} \cdot \mathrm{e}^\mu \right) \psi_x^B
 - \bar{\psi}_{x+\mathrm{a}_\mu}^B \left( \Sigma^{(2m)} \cdot
				    \mathrm{e}^\mu \right) \psi_x^A
 \nonumber \\
 && \qquad - \bar{\psi}_{x+\mathrm{a}_\mu}^A \left( \bar\Sigma^{(2m)} \cdot \mathrm{e}^\mu \right) \psi_x^B
 + \bar{\psi}_{x-\mathrm{a}_\mu}^B \left( \bar\Sigma^{(2m)} \cdot \mathrm{e}^\mu \right) \psi_x^A
\Big) \nonumber \\
 && \qquad + \bar{\psi}_x^A \left( \Sigma^{(2m)} \cdot \mathrm{e}^{2m+1} \right) \psi_x^B
 - \bar{\psi}_x^B \left( \Sigma^{(2m)} \cdot \mathrm{e}^{2m+1} \right) \psi_x^A \nonumber \\
 && \qquad - \bar{\psi}_x^A \left( \bar\Sigma^{(2m)} \cdot \mathrm{e}^{2m+1} \right) \psi_x^B
 + \bar{\psi}_x^B \left( \bar\Sigma^{(2m)} \cdot \mathrm{e}^{2m+1} \right) \psi_x^A
 \Big]
 \label{creutz_action}
\end{eqnarray}
where $\Sigma^{(2m)}$ and $\bar{\Sigma}^{(2m)}$ are $2m$-dimensional
vectors,
\begin{equation}
 \Sigma^{(2m)} = \left( \Gamma^{(2m-2)}_1, \cdots,
		  \Gamma^{(2m-2)}_{2m-1}, \id_{[2^{m-1}]} \right),
 \quad
 \bar{\Sigma}^{(2m)} = \left( \Gamma^{(2m-2)}_1, \cdots,
		  \Gamma^{(2m-2)}_{2m-1}, -\id_{[2^{m-1}]} \right),
 \label{spinor_vec2}
\end{equation}
which are similar to $\gamma^{(2m)}$ and $\bar\gamma^{(2m)}$ expressed
in (\ref{spinor_vec}), but their $2m$-th components are ``twisted''.
In this sense, this action requires ``spinor twist''.
As the case of BBTW fermion, $\mathrm{e}^\mu$ can be different from the
bond vector of the lattice.
Actualy, the lattice structure in real space is determined by the
reciprocal vectors characterizing the momentum space lattice
structure.\cite{KM:2009co}

In this case, however, the action includes non-nearest neighbor site
hoppings.
For $A$ to $B$ terms, a forward hopping is just a nearest neighbor
one, and vice versa.
On the other hand, an $A$ to $B$ backward hopping and a $B$ to $A$ forward
hopping can not be represented by a nearest, but third neighbor one (see
Fig.\ref{honeycomb_lattice}).
This means that non-nearest neighbor hoppings are required for
constructing Creutz fermion on the hyperdiamond lattice.

As the previous section, we discuss Dirac operator in momentum space
\begin{equation}
 S_{\mathrm{C}} = \int \frac{d^{2m} p}{(2\pi)^{2m}} \bar{\Psi}(-p) D(p) \Psi(p)
\end{equation}
where $\Psi$, $\bar{\Psi}$ are defined in (\ref{Dirac_spinor}), and
this Dirac operator is expanded as 
\begin{equation}
 D(p) = i \sum_{\mu=1}^{2m}\xi_\mu(p) \Gamma^{(2m)}_\mu  \label{Dirac_op3}
\end{equation}
with
\begin{eqnarray}
 \xi_\mu(p) &=& \left( \prod_{\nu=\mu+1}^{2m} s_\nu \right) \left( c_\mu
	       \sum_{\nu=1}^\mu \sin p_\nu + \sin p_{\mu+1} \right)
 \quad \mbox{for} \quad \mu = 1, \cdots, 2m-1, \nonumber \\ \label{Creutz_coefficient1}
\\
 \xi_{2m}(p) &=& B \left( C + c_{2m}\sum_{\nu=1}^{2m}\cos p_\nu \right).\label{Creutz_coefficient2}
\end{eqnarray}
Immediately two poles of this operator are obtained as $p^{(\pm)} = \pm
\left( \tilde{p}_{\mathrm{C}}, \cdots, \tilde{p}_{\mathrm{C}}\right)$ with $\cos \tilde{p}_{\mathrm{C}} = C$.
To prohibit extra poles such as $p = \left( \tilde{p}_{\mathrm{C}}, \cdots,
\tilde{p}_{\mathrm{C}}, \pi - \tilde{p}_{\mathrm{C}} \right)$, the range
of the parameter $C$ 
should be
\begin{equation}
 \frac{m-1}{m} < C < 1. \label{minimal_cond}
\end{equation}
This minimal-doubling condition reduces to $1/2<C<1$ in the $d=4$ $(m=2)$ case, as is consistent with the result in Ref.~\citen{Creutz:2007af}.
In addition, in the case $C=1$, the number of poles is one, but this only one pole
corresponds to an unphysical fermion as in the four-dimensional case \cite{KM:2009co}. 
Thus, it is impossible to define Creutz fermion on the exact hyperdiamond lattice in general even dimensions since the bond vectors (\ref{bond_dist}) reduce to these of an exact hyperdiamond lattice only with the parameters $B=C=1$.
As a result, the discrete symmetry of the lattice itself and the action
breaks down from symmetric group $\mathfrak{S}_{d+1}$ to
$\mathfrak{S}_{d}$.
That means that the action is not invariant under permutations of $d+1$
but $d$ bonds of the hyperdiamond lattice.
This situation is also the case as four-dimensional Creutz fermion.
We note that the Dirac operator of this lattice fermion can be written in the same form as Eq.(\ref{chiralrep}), and thus it is easily confirmed that the action is chiral-invariant.

In Ref.\ \citen{Creutz:2007af}, Creutz chose the two parameters to place
poles on the four-dimensional hyperdiamond lattice.
To discuss a lattice structure of poles in momentum space, we then expand Dirac
operator (\ref{Dirac_op3}) around a pole with $p_\mu = \tilde{p}_{\mathrm{C}} + q_\mu$,
\begin{eqnarray}
 \xi_\mu(p) &=& C \left( \prod_{\nu=\mu+1}^{2m} s_\nu \right) \left( c_\mu
		 \sum_{\nu=1}^\mu q_\nu + q_{\mu+1} \right) + {\cal
 O}(q^2) \quad \mbox{for} \quad \mu = 1, \cdots, 2m-1, \nonumber \\ \label{Creutz_coefficient3}
\\
 \xi_{2m}(p) &=& -BS c_{2m} \sum_{\nu=1}^{2m} q_\nu + {\cal O}(q^2)\label{Creutz_coefficient4}
\end{eqnarray}
with $S = \sqrt{1-C^2}$.
In the case $C=1$, since the coefficient of $\Gamma^{(2m)}_{2m}$
vanishes, the Dirac operator (\ref{Dirac_op3}) can not be reduced to
$2m$-dimensional covariant form, while $(2m-1)$-dimensional covariance is
preserved,
\begin{equation}
 D(p) = i \sum_{\mu=1}^{2m-1} \xi_\mu(p) \Gamma^{(2m)}_\mu
\equiv i \slash{k}. \label{Dirac_op4}
\end{equation}
This means that only one $(2m-1)$-dimensional chiral fermion is derived from
$2m$-dimensional lattice.
Although this is similar to domain-wall
fermion\cite{Kaplan:1992bt,Furman:1994ky}, this fermion
(\ref{Dirac_op4}) is not localized in $(2m-1)$-dimensional subspace. 
Domain-wall fermion is obtained by twisting boundary condition of the
mass term in fifth direction to be localized on four-dimensional
subspace.
The reason why this kind of excitation appears in Creutz fermion is
explained as follows.
In this case (\ref{Dirac_op4}), the gamma matrix for $2m$-direction is
treated specially in (\ref{spinor_vec2}) because its sign is twisted.
In fact, the reduced dimension corresponds to the direction in which the
gamma matrix is twisted.
However the domain-wall-like fermion (\ref{Dirac_op4}) is not localized on
$(2m-1)$-dimensional subspace because we do not consider the mass term
depending on the $2m$-th direction.

We then consider momentum space structure.
According to the coefficients of gamma matrices, momentum space basis is
obtained as
\begin{equation}
 \begin{array}{lccrrrrrrrc}
 \mathrm{b}_1 & = & (& Cc_1 s_2 \cdots s_{d-1} s_d ,& Cc_2 s_3 \cdots s_d ,& \cdots & , & Cc_{d-2} s_{d-1} s_d, & Cc_{d-1} s_d,& -BSc_d &) \\
 \mathrm{b}_2 & = & (& Cs_2 \cdots s_{d-1} s_d ,& Cc_2 s_3 \cdots s_d ,& \cdots & , & Cc_{d-2} s_{d-1} s_d, & Cc_{d-1} s_d,& -BSc_d &) \\
 \mathrm{b}_3 & = & (& 0 ,& Cs_3 \cdots s_d ,& \cdots & , & Cc_{d-2} s_{d-1} s_d, & Cc_{d-1} s_d,& -BSc_d &) \\
  &\vdots &&&&&&&& \\
 \mathrm{b}_{d-1} & = & (& 0,& 0,& \cdots & , & Cs_{d-1} s_d , & Cc_{d-1} s_d,& -BSc_d & ) \\  
 \mathrm{b}_{d} & = & (& 0,& 0,& \cdots & , & 0, & Cs_d ,& -BSc_d & ) \\  
 \end{array} . \label{dual_vec}
\end{equation}
Thus, an angle between these vectors is derived from a norm and an inner
product,
\begin{equation}
 \cos \eta = \frac{\mathrm{b}_\mu \cdot
  \mathrm{b}_\nu}{\left|\mathrm{b}_\mu\right|\left|\mathrm{b}_\nu\right|} = \frac{
  B^2 S^2 c_d^2 + C^2 s_d^2 c_{d-1}}{ B^2 S^2 c_d^2 + C^2 s_d^2}.
\end{equation}

We then consider the condition that the lattice of poles in momentum space becomes the exact hyperdiamond lattice, which we call the hyperdiamond condition (Creutz condition\cite{Creutz:2007af}).
This is given by $\cos \eta = 1/2$ as compared with (\ref{primitive_vec}).
Another condition that the pole lattice becomes orthogonal, which we call the orthogonal
condition (Bori\c{c}i condition\cite{Borici:2007kz}) is given by $\cos \eta = 0$.
The former, hyperdiamond condition requires an additional condition that $d+1$
distances coincide, $\left|p^{(+)} - p^{(-)}\right| = \left|p^{(+)} -
\left( p^{(-)} + 2\pi \mathrm{b}_1\right)\right| = \cdots = \left|p^{(+)} -
\left( p^{(-)} + 2\pi \mathrm{b}_{d}\right)\right|$.
As a result, we obtain Creutz condition
\begin{equation}
 \tilde{p}_{\mathrm{C}} = \frac{\pi}{d+1}, \qquad 
 C = \cos \left( \frac{\pi}{d+1} \right), \quad
 B = (d+1) \cot \left( \frac{\pi}{d+1} \right), \label{Creutz_cond}
\end{equation}
and Bori\c{c}i condition
\begin{equation}
 C = \cos \tilde p_{\mathrm{C}} , \quad
 B = \sqrt{d+1} \cot \tilde p_{\mathrm{C}}. \label{Borici_cond}
\end{equation}
Here we can choose arbitrary $\tilde{p}_{\mathrm{C}}$ satisfying the
minimal-doubling condition (\ref{minimal_cond}).
These conditions are reduced to the four-dimensional version in Ref.\ \citen{Creutz:2007af} by choosing $d=4$ and replacing $B \to \sqrt{d+1}B$.

We now remark consistency between the Creutz condition
(\ref{Creutz_cond}) and the minimal-doubling condition (\ref{minimal_cond}).
The parameter should behave as $C \to 1$ with the large $d$ limit
because the lower bound of (\ref{minimal_cond}) goes to its upper bound
as $\mathcal{O}(1/d)$.
Indeed the Creutz condition gives $C \to 1$ with $d
\gg 1$, but it converges faster as $\mathcal{O}(1/d^2)$ since $\cos (1/x) \sim 1 - 1/(2x^2)$ with large $x$.
Thus the Creutz condition (\ref{Creutz_cond}) satisfies the
minimal-doubling condition (\ref{minimal_cond}) for any dimensions.

On the other hand, for the Bori\c{c}i condition (\ref{Borici_cond}), 
in particular $C=S=1/\sqrt{2}$ is chosen for simplicity in Ref.\
\citen{Borici:2007kz}. 
This parameters, however, can not satisfy the
condition (\ref{minimal_cond}) with sufficiently large $d$.
Generally speaking, the minimal-doubling condition (\ref{minimal_cond}) hardly realizes in the
large $d$ limit since the range of the parameter becomes so narrow.

\subsection*{Some related lattice actions}

Some lattice actions which are related to four-dimensional Creutz action have been considered in \citen{KM:2009co} to discuss remarkable features of Creutz action: (i) non-nearest neighbor hoppings, (ii) twisting spinor structure, and (iii) distorting the hyperdiamond lattice.
It is shown that the lattice action based on the hyperdiamond lattice which includes only physical lattice fermions satisfies all of the above conditions, and unphysical fermions arise in the other lattice actions.
In particular, the non-nearest neighbor hopping is essential for constructing the physical lattice fermion.
Both of $i\Gamma$-terms and $\Gamma\Gamma_{d+1}$-terms arise in lattice
actions without non-nearest neighbor hoppings, and thus
Nielsen-Ninomiya's no-go theorem, which is based on Poincar\'e-Hopf
theorem for vector function, cannot be applied to this kind of lattice
actions.
However, adding non-nearest neighbor hoppings to obtain Dirac operator including only $i\Gamma$-terms, the lattice action lacks the high discrete symmetry of
the hyperdiamond lattice due
to non-nearest neighbor hoppings.%
\footnote{
The lattice action based on the distorted hyperdiamond lattice lacks the sufficient discrete symmetry
even if one drops the non-nearest hopping terms because, in the first place, the lattice distortion lowers the discrete symmetry of the lattice. 
}
The sufficient discrete symmetry is required for suppressing the
redundant operators generated by the loop corrections, and thus it is an
important property for the lattice simulation.

We now consider some higher-dimensional lattice actions from the viewpoint of the three features of Creutz action discussed above.
One of related actions is obtained by dropping the non-nearest neighbor hopping terms of Creutz action, which was originally considered in \citen{Bedaque:2008jm} and called Dropped Creutz action\cite{KM:2009co}.
Its higher-dimensional generalization is easily constructed on the hyperdiamond lattice from (\ref{creutz_action}),
\begin{eqnarray}
  S_{\mathrm{dC}} & = & \frac{1}{2} \sum_x \Big[\sum_{\mu=1}^{2m}  \Big(
  \bar{\psi}_{x-\mathrm{a}_\mu}^A \left( \Sigma^{(2m)} \cdot \mathrm{e}^\mu \right) \psi_x^B
 - \bar{\psi}_{x+\mathrm{a}_\mu}^B \left( \Sigma^{(2m)} \cdot
				    \mathrm{e}^\mu \right) \psi_x^A
 \nonumber \\
 && \qquad + \bar{\psi}_x^A \left( \Sigma^{(2m)} \cdot \mathrm{e}^{2m+1} \right) \psi_x^B
 - \bar{\psi}_x^B \left( \Sigma^{(2m)} \cdot \mathrm{e}^{2m+1} \right) \psi_x^A \Big].
 \label{d_creutz_action}
\end{eqnarray}
This action is similar to BBTW action (\ref{BBTW_action}) and given by the modification of $2m$-th gamma matrix, $i\Gamma_{2m}^{(2m)} \to \Gamma_{2m+1}^{(2m)}\Gamma_{2m}^{(2m)}$.
It also possesses a pole at $p=0$, but its covariance is broken as $i \vec{\Gamma}^{(2m)}\cdot \vec{k} + \Gamma_{2m+1}^{(2m)}\Gamma_{2m}^{(2m)} k_{2m}$.

Another lattice action based on the (distorted) hyperdiamond lattice is
called Untwisted Creutz action\cite{KM:2009co}, which is given by
modifying the spinor structure of Creutz action as $\Sigma^{(2m)} \to
\gamma^{(2m)}$.
The higher-dimensional generalization is given by
\begin{eqnarray}
  S_{\mathrm{utC}} & = & \frac{1}{2} \sum_x \Big[\sum_{\mu=1}^{2m}  \Big(
  \bar{\psi}_{x-\mathrm{a}_\mu}^A \left( \gamma^{(2m)} \cdot \mathrm{e}^\mu \right) \psi_x^B
 - \bar{\psi}_{x+\mathrm{a}_\mu}^B \left( \gamma^{(2m)} \cdot
				    \mathrm{e}^\mu \right) \psi_x^A
 \nonumber \\
 && \qquad - \bar{\psi}_{x+\mathrm{a}_\mu}^A \left( \bar\gamma^{(2m)} \cdot \mathrm{e}^\mu \right) \psi_x^B
 + \bar{\psi}_{x-\mathrm{a}_\mu}^B \left( \bar\gamma^{(2m)} \cdot \mathrm{e}^\mu \right) \psi_x^A
\Big) \nonumber \\
 && \qquad + \bar{\psi}_x^A \left( \gamma^{(2m)} \cdot \mathrm{e}^{2m+1} \right) \psi_x^B
 - \bar{\psi}_x^B \left( \gamma^{(2m)} \cdot \mathrm{e}^{2m+1} \right) \psi_x^A \nonumber \\
 && \qquad - \bar{\psi}_x^A \left( \bar\gamma^{(2m)} \cdot \mathrm{e}^{2m+1} \right) \psi_x^B
 + \bar{\psi}_x^B \left( \bar\gamma^{(2m)} \cdot \mathrm{e}^{2m+1} \right) \psi_x^A
 \Big].
 \label{ut_creutz_action}
\end{eqnarray}
This action includes the minimal-doubling fermions, but their covariance are also broken as $i \vec{\Gamma}^{(2m)}\cdot \vec{k} + \Gamma_{2m}^{(2m)} k_{2m}$.
In the case $C=1$, they are reduced to only one, but unphysical pole $i \vec{\Gamma}^{(2m)}\cdot \vec{k}$ which is the same as (\ref{Dirac_op4}).

As a result, the higher-dimensional hyperdiamond lattice actions including a physical pole is Creutz and BBTW action.
However BBTW action also includes unphysical poles. 
Thus the lattice actions which yield only physical poles are Creutz action and the modified Creutz action which we present in the next section, called Appended Creutz action.
This situation is the same as the four-dimensional case\cite{KM:2009co}.

\section{Higher-dimensional Appended Creutz action}\label{sec:appended}

It has been pointed out in \citen{Bedaque:2008jm,KM:2009co} that
Creutz action includes not only nearest neighbor but non-nearest
neighbor hopping terms.
These non-nearest interaction terms represent hoppings to non-nearest
sites, but to nearest unit cells.
Although the locality of the continuum theory is often broken via non-nearest
hopping terms of the lattice field theory, the locality of Creutz action
is not broken in the continuum limit\cite{Cichy:2008gk,Cichy:2008nt}
because the action includes only hopping terms to nearest unit cells.

\begin{figure}[tbp]
\begin{center}
 \includegraphics[width=9.3em]{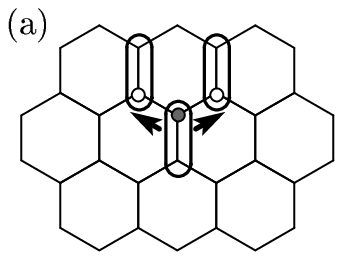} \qquad
 \includegraphics[width=9.3em]{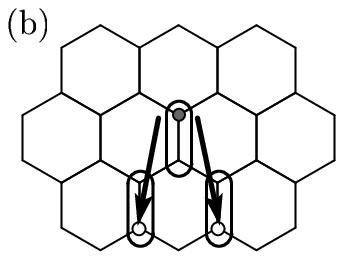} \qquad
 \includegraphics[width=9.3em]{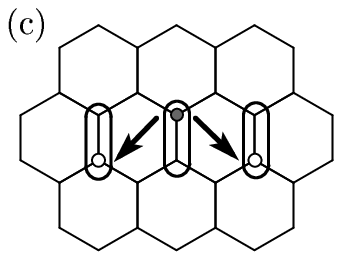}
\end{center}
\caption{Two-dimensional analogues of hoppings to nearest neighbor unit
 cells
: (a) nearest site hoppings, (b) non-nearest site hoppings included by Creutz action and (c) the remaining nearest unit cell hoppings included only by Appended Creutz action.}
\label{honeycomb}
\end{figure}

On the other hand, the number of nearest unit cells is $d(d+1)$ as
explained in section \ref{sec:hyperdiamond} , and thus all of nearest
unit cell hoppings are not yet included by Creutz action.
We then consider Creutz action including all of nearest unit cell
hopping terms,
\begin{eqnarray}
  S_{\mathrm{aC}} & = & S_{\mathrm{C}} + \frac{1}{2} \sum_x \Big[ \sum_{\mu < \nu} \Big( \bar\psi_{x - \mathrm{a}_\mu +
			 \mathrm{a}_\nu}^A \left( \Sigma^{(2m)} \cdot \left(
								\mathrm{e}_\mu
							       -
							       \mathrm{e}_\nu
								      \right)
					    \right) \psi_x^B - \bar\psi_{x + \mathrm{a}_\mu -
			 \mathrm{a}_\nu}^B \left( \Sigma^{(2m)} \cdot \left(
								\mathrm{e}_\mu
							       -
							       \mathrm{e}_\nu
								      \right)
					    \right) \psi_x^A
			 \nonumber \\
 && \qquad - \bar\psi_{x + \mathrm{a}_\mu -
			 \mathrm{a}_\nu}^A \left( \bar \Sigma^{(2m)} \cdot \left(
								\mathrm{e}_\mu
							       -
							       \mathrm{e}_\nu
								      \right)
					    \right) \psi_x^B + \bar\psi_{x - \mathrm{a}_\mu +
			 \mathrm{a}_\nu}^B \left( \bar\Sigma^{(2m)} \cdot \left(
								\mathrm{e}_\mu
							       -
							       \mathrm{e}_\nu
								      \right)
					    \right) \psi_x^A
\Big) \Big]. \label{appended_Creutz}
\end{eqnarray}
The additive terms of (\ref{appended_Creutz}) represent the remaining
nearest unit cell hoppings shown in Fig. \ref{honeycomb}, and this
action is a higher-dimensional generalization of Appended Creutz
action presented in our previous work\cite{KM:2009co}.

The momentum space representation of the Dirac operator is given by 
\begin{equation}
 D(p) = i \sum_{\mu = 1}^{2m} \left( \xi_\mu(p) + \xi_\mu^{(+)}(p)
			      \right) \Gamma_\mu^{(2m)}
\end{equation}
where $\xi_\mu(p)$ are already defined as (\ref{Creutz_coefficient1}) and
(\ref{Creutz_coefficient2}), and $\xi_\mu^{(+)}(p)$ are additive terms
derived from the remaining nearest unit cell terms,
\begin{eqnarray}
 \xi_\mu^{(+)}(p) & = & \left( \prod_{\nu = \mu+1}^{2m} s_\nu \right) \left(
			 (c_\mu - 1)\sum_{\nu=1}^\mu \sin(p_\nu -
			 p_{\mu+1}) + c_\mu \sum_{\nu = 1}^{\mu-1}\sum_{\lambda = \mu + 2}^{2m} \sin
			 (p_\nu - p_\lambda) \right. \nonumber \\
 && \left. + \sum_{\nu = \mu+2}^{2m}
			 \sin(p_{\mu+1}-p_\nu) \right) \qquad \mbox{for} \quad \mu = 1, \cdots, 2m - 1, \\
 \xi_{2m}^{(+)}(p) & = & 0.
\end{eqnarray}
Thus we obtain exactly the same two poles as Creutz action, $p^{(\pm)} = \pm (\tilde
p_\mathrm{C}, \cdots, \tilde p_\mathrm{C})$ with $\cos \tilde p_\mathrm{C} = C$, and the minimal-doubling condition is also given by (\ref{minimal_cond}).

We then expand the Dirac operator around a pole as $p_\mu = \tilde p_{\mathrm{C}} + q_\mu$ to investigate the momentum space structure,
\begin{eqnarray}
 \xi_\mu^{(+)}(p) & = & 2m \left(\prod_{\nu=\mu+1}^{2m}s_\nu\right)
  \left( c_\mu \sum_{\nu=1}^{\mu} q_\nu + q_{\mu+1}\right) + {\cal
  O}(q^2)
 \quad \mbox{for} \quad \mu = 1, \cdots, 2m - 1. \nonumber \\
\end{eqnarray}
This expression is proportional to (\ref{Creutz_coefficient3}), and thus the total contribution to the coefficient of $\Gamma_\mu^{(2m)}$ becomes $\xi_\mu(p) + \xi_\mu^{(+)}(p) = (1 + d/C) \xi_\mu(p)$.
Thus the momentum space basis is obtained,
\begin{equation}
 \begin{array}{lccrrrrrrc}
 \mathrm{b}_1 & = & (& (d+C)c_1 s_2 \cdots s_{d-1} s_d ,& (d+C)c_2 s_3 \cdots s_d ,& \cdots & , & (d+C)c_{d-1} s_d,& -BSc_d &) \\
 \mathrm{b}_2 & = & (& (d+C)s_2 \cdots s_{d-1} s_d ,& (d+C)c_2 s_3 \cdots s_d ,& \cdots & , & (d+C)c_{d-1} s_d,& -BSc_d &) \\
 \mathrm{b}_3 & = & (& 0 ,& (d+C)s_3 \cdots s_d ,& \cdots & , & (d+C)c_{d-1} s_d,& -BSc_d &) \\
  &\vdots &&&&&&& \\
 \mathrm{b}_{d-1} & = & (& 0,& 0,& \cdots & , & (d+C)c_{d-1} s_d,& -BSc_d & ) \\  
 \mathrm{b}_{d} & = & (& 0,& 0,& \cdots & , & (d+C)s_d ,& -BSc_d & ) \\  
 \end{array} . \label{dual_vec_append}
\end{equation}
As the case of Creutz action discussed in section \ref{sec:creutz}, the angle between the reciprocal vectors (\ref{dual_vec_append}) is given by
\begin{equation}
 \cos \eta = \frac{\mathrm{b}_\mu \cdot
  \mathrm{b}_\nu}{\left|\mathrm{b}_\mu\right|\left|\mathrm{b}_\nu\right|} = \frac{
  B^2 S^2 c_d^2 + (d+C)^2 s_d^2 c_{d-1}}{ B^2 S^2 c_d^2 + (d+C)^2 s_d^2}.
\end{equation}
Therefore, we obtain Creutz condition such that two poles constitute the
exact hyperdiamond lattice on the momentum space, 
\begin{equation}
 \tilde{p}_{\mathrm{C}} = \frac{\pi}{d+1}, \qquad 
 C = \cos \left( \frac{\pi}{d+1} \right), \quad
 B = (d+1) \frac{d+\cos\left( \frac{\pi}{d+1} \right)}{\sin \left(
			\frac{\pi}{d+1} \right)}, \label{appended_Creutz_cond}
\end{equation}
and Bori\c{c}i condition such that reciprocal vectors are orthogonal,
\begin{equation}
 C = \cos \tilde p_{\mathrm{C}} , \quad
 B = \sqrt{d+1} \frac{d + \cos \tilde p_{\mathrm{C}}}{\sin \tilde
 p_{\mathrm{C}}}.
\end{equation}
These conditions are almost the same as (\ref{Creutz_cond}) and
(\ref{Borici_cond}).
They are easily given by shifting the parameter $B \propto C/S \to (d + C)/S$ while $C$ is not modified.
Thus the modified Creutz condition (\ref{appended_Creutz_cond}) also satisfies the minimal-doubling condition (\ref{minimal_cond}) for any dimensions.

In the case of Appended Creutz action, it is shown that the same
minimal-doubling poles and a slightly modified momentum space structure
are obtained as the case of Creutz action.
In the action, $2m$-dimensional vectors $\mathrm{e}_\mu -
\mathrm{e}_\nu$ are considered in hopping terms, but $2m$-th components of
them are constantly zero.
Thus these terms hardly affect the lattice action in $2m$-th direction.
As a result, we conclude that the effect of the remaining nearest unit
cell hopping terms is small.

\section{Discussion}\label{sec:summary}

In this paper we generalize the several types of fermions on the hyperdiamond lattice to higher even dimensions and extract the characteristics of them, some of which are common to the four-dimensional case and others of which are specific to the higher-dimensional cases.

In Sec. \ref{sec:BBTW}, it is pointed out that BBTW fermions in higher even dimensions inevitably produce unphysical poles of fermion propagator since the BBTW Dirac operator independently includes $i\Gamma$-terms and $\Gamma\Gamma_{d+1}$-terms as the four-dimensional case\cite{KM:2009co}.
Thus we can conclude that simple construction of Dirac fermion on higher-dimensional hyperdiamond lattice leads to unphysical degrees of freedom.

In Sec. \ref{sec:creutz} we also construct the minimal-doubling fermion action called Creutz action in general even dimensions.
We show that it is defined on the distorted lattice as long as one keeps physicality of the poles of propagator, and the action loses the high discrete symmetry of the original lattice as the four-dimensional case. 
It is pointed out that the range of the parameters for Creutz fermion to yield the minimal amount of physical fermions becomes narrower with the dimension getting higher.
Thus, in higher dimensions it becomes more and more difficult to realize minimal-doubling in Creutz fermion.
We also derive the higher-dimensional versions of the specific parameter conditions, Creutz condition and Bori\c{c}i condition.
These generalized conditions reduce to the original four-dimensional conditions by setting $d=4$.
In addition we generalize the subspecies of Creutz and BBTW actions discussed in our recent paper.
Then it is shown that all of them include unphysical degrees of freedom as the four-dimensional case.

In Sec. \ref{sec:appended} we study a higher dimensional version of the new minimal-doubling fermion, Appended Creutz action.
Then it is pointed out that this fermion has the modified momentum-space structure leading to the modified Creutz and Bori\c{c}i conditions compared to the original Creutz action.  
This type of fermion has been discussed for the first time in our recent paper, so it will be valuable to investigate the properties in more detail.

Finally, let us comment on symmetries of BBTW and Creutz fermions.
BBTW fermion action in general even dimensions has high discrete symmetry of alternating group $\mathfrak{A}_{d+1}$ as well as chiral symmetry.
However these symmetries have nothing to do with physics since this action yields unphysical degrees of freedom in principle.
On the other hand Creutz action in general even dimensions describes only physical fermions, thus it is meaningful to discuss symmetries of this action.
It has chiral symmetry, but possesses only the discrete symmetry of $\mathfrak{S}_{d}$.
As seen in the four-dimensional case\cite{Bedaque:2008xs}, the low discrete symmetry may lead to the generation of marginal and relevant operators through quantum corrections.
We need to look into this topic in the future work.

\section*{Acknowledgments}
We would like to thank T. Onogi for reading the manuscript and useful discussions.
We also thank Y. Kikukawa for valuable comments.
TM is supported by Grant-in-Aid for the Japan Society for Promotion o Science (JSPS) Research Fellows.

%

  


\begin{thebibliography}{10}

\bibitem{neto:109}
A.~H.~C. Neto, F.~Guinea, N.~M.~R. Peres, K.~S. Novoselov, and A.~K. Geim, {\it
  {The electronic properties of graphene}},  {\em Rev. Mod. Phys.} {\bf 81}
  (2009) 109, [\href{http://arxiv.org/abs/0709.1163}{{\tt arXiv:0709.1163}}].

\bibitem{Nielsen:1980rz}
H.~B. Nielsen and M.~Ninomiya, {\it {Absence of Neutrinos on a Lattice. 1.
  Proof by Homotopy Theory}},  {\em Nucl. Phys.} {\bf B185} (1981) 20.

\bibitem{Nielsen:1981xu}
H.~B. Nielsen and M.~Ninomiya, {\it {Absence of Neutrinos on a Lattice. 2.
  Intuitive Topological Proof}},  {\em Nucl. Phys.} {\bf B193} (1981) 173.

\bibitem{Nielsen:1981hk}
H.~B. Nielsen and M.~Ninomiya, {\it {No Go Theorem for Regularizing Chiral
  Fermions}},  {\em Phys. Lett.} {\bf B105} (1981) 219.

\bibitem{Kaplan:1992bt}
D.~B. Kaplan, {\it {A Method for simulating chiral fermions on the lattice}},
  {\em Phys. Lett.} {\bf B288} (1992) 342,
  [\href{http://arxiv.org/abs/hep-lat/9206013}{{\tt hep-lat/9206013}}].

\bibitem{Furman:1994ky}
V.~Furman and Y.~Shamir, {\it {Axial symmetries in lattice QCD with Kaplan
  fermions}},  {\em Nucl. Phys.} {\bf B439} (1995) 54,
  [\href{http://arxiv.org/abs/hep-lat/9405004}{{\tt hep-lat/9405004}}].

\bibitem{Neuberger:1998wv}
H.~Neuberger, {\it {More about exactly massless quarks on the lattice}},  {\em
  Phys. Lett.} {\bf B427} (1998) 353,
  [\href{http://arxiv.org/abs/hep-lat/9801031}{{\tt hep-lat/9801031}}].

\bibitem{Ginsparg:1981bj}
P.~H. Ginsparg and K.~G. Wilson, {\it {A Remnant of Chiral Symmetry on the
  Lattice}},  {\em Phys. Rev.} {\bf D25} (1982) 2649.

\bibitem{Wilczek:1987kw}
F.~Wilczek, {\it {Lattice Fermions}},  {\em Phys. Rev. Lett.} {\bf 59} (1987)
  2397.

\bibitem{Karsten:1981gd}
L.~H. Karsten, {\it {Lattice fermions in euclidean space-time}},  {\em Phys.
  Lett.} {\bf B104} (1981) 315.

\bibitem{Bedaque:2008jm}
P.~F. Bedaque, M.~I. Buchoff, B.~C. Tiburzi, and A.~Walker-Loud, {\it {Search
  for Fermion Actions on Hyperdiamond Lattices}},  {\em Phys. Rev.} {\bf D78}
  (2008) 017502, [\href{http://arxiv.org/abs/0804.1145}{{\tt
  arXiv:0804.1145}}].

\bibitem{KM:2009co}
T.~Kimura and T.~Misumi, {\it {Characters of Lattice Fermions Based on the
  Hyperdiamond Lattice}},  \href{http://arxiv.org/abs/0907.1371}{{\tt
  arXiv:0907.1371}}.

\bibitem{Creutz:2007af}
M.~Creutz, {\it {Four-dimensional graphene and chiral fermions}},  {\em JHEP}
  {\bf 0804} (2008) 017, [\href{http://arxiv.org/abs/0712.1201}{{\tt
  arXiv:0712.1201}}].

\bibitem{Bedaque:2008xs}
P.~F. Bedaque, M.~I. Buchoff, B.~C. Tiburzi, and A.~Walker-Loud, {\it {Broken
  Symmetries from Minimally Doubled Fermions}},  {\em Phys. Lett.} {\bf B662}
  (2008) 449, [\href{http://arxiv.org/abs/0801.3361}{{\tt arXiv:0801.3361}}].

\bibitem{Capitani:2009yn}
S.~Capitani, J.~Weber, and H.~Wittig, {\it {Minimally doubled fermions at one
  loop}},  {\em Phys. Lett.} {\bf B681} (2009) 105,
  [\href{http://arxiv.org/abs/0907.2825}{{\tt arXiv:0907.2825}}].

\bibitem{Susskind:1976jm}
L.~Susskind, {\it {Lattice Fermions}},  {\em Phys. Rev.} {\bf D16} (1977) 3031.

\bibitem{Celmaster:1982ht}
W.~Celmaster, {\it {Gauge Theories on the Body - Centerd Hypercubic Lattice}},
  {\em Phys. Rev.} {\bf D26} (1982) 2955.

\bibitem{Celmaster:1983jq}
W.~Celmaster and F.~Krausz, {\it {Fermion Mutilation on a Body centered
  Tesseract}},  {\em Phys. Rev.} {\bf D28} (1983) 1527.

\bibitem{Drouffe:1983kq}
J.~M. Drouffe and K.~J.~M. Moriarty, {\it {Gauge Theories on a Simplical
  Lattice}},  {\em Nucl. Phys.} {\bf B220} (1983) 253.

\bibitem{Borici:2007kz}
A.~Bori\c{c}i, {\it {Creutz fermions on an orthogonal lattice}},  {\em Phys.
  Rev.} {\bf D78} (2008) 074504, [\href{http://arxiv.org/abs/0712.4401}{{\tt
  arXiv:0712.4401}}].

\bibitem{Cichy:2008gk}
K.~Cichy, J.~Gonzalez~Lopez, K.~Jansen, A.~Kujawa, and A.~Shindler, {\it
  {Twisted Mass, Overlap and Creutz Fermions: Cut-off Effects at Tree-level of
  Perturbation Theory}},  {\em Nucl. Phys.} {\bf B800} (2008) 94,
  [\href{http://arxiv.org/abs/0802.3637}{{\tt arXiv:0802.3637}}].

\bibitem{Cichy:2008nt}
K.~Cichy, J.~Gonzalez~Lopez, and A.~Kujawa, {\it {A comparison of the cut-off
  effects for Twisted Mass, Overlap and Creutz fermions at tree-level of
  Perturbation Theory}},  {\em Acta Phys. Polon.} {\bf B39} (2008) 3463,
  [\href{http://arxiv.org/abs/0811.0572}{{\tt arXiv:0811.0572}}].

\end{thebibliography}

\providecommand{\href}[2]{#2}\begingroup\raggedright\endgroup

\end{document}